\documentclass{osa-article}
\usepackage{float}
\usepackage{mwe}

\journal{oe}

\begin{document}

\title{Molecular Quantum Wakes for Clearing Fog}

\author{Malte C. Schroeder,\authormark{1} Ilia Larkin,\authormark{2} Thomas Produit,\authormark{1} Eric W. Rosenthal,\authormark{3} Howard Milchberg\authormark{2} and Jean-Pierre Wolf\authormark{1,*}}

\address{\authormark{1}Dept. of Applied Physics, University of Geneva, Chemin de Pinchat 22, 1211 Geneva 4, Switzerland\\
\authormark{2}Institute for Research in Electronics and Applied Physics, University of Maryland, College Park, MD 20742, USA\\
\authormark{3}U.S. Naval Research Laboratory, Washington, D.C. 20375, USA\\}

\email{\authormark{*}Jean-Pierre.Wolf@unige.ch} 



\begin{abstract}
High intensity laser filamentation in air has recently demonstrated that, through plasma generation and its associated shockwave, fog can be cleared around the beam, leaving an optically transparent path to transmit light. However, for practical applications like free-space optical communication (FSO), channels of multi-centimeter diameters over kilometer ranges are required, which is extremely challenging for a plasma based method. Here we report a radically different approach, based on quantum control. We demonstrate that fog clearing can also be achieved by producing molecular quantum wakes in air, and that neither plasma generation nor filamentation are required. The effect is clearly associated with the rephasing time of the rotational wave packet in N$_2$. Pump excitation provided in the form of resonant trains of 8 pulses separated by the revival time are able to transmit optical data through fog with initial extinction as much as $-6$\,dB.
\end{abstract}

\section{Introduction}
Light scattering by small particles is a key issue for several optical applications, such as imaging through turbid media or free space optical telecommunication (FSO) between ground based stations and satellites. Many approaches based on adaptive optics and spatio-temporal shaping of laser beams have been proposed for improving the propagation through scattering media and for avoiding speckle patterns formation \cite{Katz2011, Mosk2012, Katz2014, McCabe2011, Vellekoop2007, Boniface2017}. These methods, however, are of limited use for dense and extended structures such as atmospheric fog and clouds. For free space optical networks, this critical issue is currently mitigated by the multiplication of transmitting/receiving stations, which increases complexity and cost. Until recently no active method for opening clear optical transmission channels through clouds or fog was available. The propagation of ultrashort and high intensity lasers through the atmosphere provided a radically new concept, based on plasma generation inherent to laser filamentation \cite{Berge2007, Schimmel2018, DeLaCruz2016}. Laser filaments are self-sustained light structures with a typical core diameter $d_{core}\sim 100\,\mu$m (at $800$\,nm) and up to hundreds of meters in length, orders of magnitude beyond the linear diffraction limit. They stem from the dynamic balance between self-focusing (Kerr effect) and defocusing (diffraction, plasma-induced refraction, etc.) that takes place during non-linear laser propagation in air \cite{Couairon2002}. Filaments bear high intensities ($10-100$\,TW/cm$^2$) and generate a low density plasma during propagation. The plasma recombination thermalizes quickly, resulting in a very rapid local increase of pressure that drives a single cycle cylindrical acoustic wave \cite{Jhajj2014, Wahlstrand2014, Cheng2013, Lahav2014, Point2015}. Over a time window of $0.1-1$\,ms, this leads to a reduced air density channel significantly wider than the filament core \cite{Cheng2013, Lahav2014, Point2015, Vidal2000, Yu2003}. When launched in fog (water droplet diameter centered around $5$\,$\mu$m), this filament-induced shockwave radially ejects the scatterers out of the beam \cite{Schimmel2018, DeLaCruz2016}. This fog clearing mechanism requires modest energy deposition, of the order of less than $1$\,mJ/m \cite{Rosenthal2018, Houard2016}. For droplets densities $N_d \sim 10^5\,$cm$^{-3}$, it was shown that the filament formation is not significantly altered \cite{Courvoisier2003}. This is the case in the vast majority of scenarios in the field \cite{Courvoisier2003, Skupin2004, Kolesik2004}, and is the case for our fog chamber as discussed below. Under these conditions, there is little additional energy cost due to this droplet clearing mechanism, as the energy deposition and acoustic wave generation through plasma generation by the filament occurs in any case as a consequence of its propagation.\par

However, applying the filament-induced clearing method to real-scale FSO may be challenging. Opened communication channels require lengths of at least several hundred meters and diameters of $\sim10$\,cm. Laboratory scale experiments demonstrated that filament-induced acoustic waves could clear regions exceeding the filament dimensions, but not larger than $0.5-1$\,cm \cite{DeLaCruz2016}. Engineering the laser pulse properties (adaptive optics and temporal focusing) in order to achieve uniform filament excitation over several hundred meters appears particularly challenging.\par
In the present paper, we generate and coherently control rotational wave packets in the air molecules (in particular N$_2$) to produce an acoustic wave to open transmission channels in fog and clouds over short distances without requiring plasma formation or filamentation. Larger diameter beams and/or mid-IR lasers could then be used for fog clearing, which when filamenting generate lower density plasma unsuited for shockwave creation, but for which long distance propagation is more easily achievable and controllable \cite{Panagiotopoulos2015,Tochitsky2019}.

\section{Coherent control of rotational wave packets in air molecules}

The effects of laser induced molecular alignment and field-free coherent rotation on laser propagation have been investigated by several authors \cite{Calegari2008, Renard2005, Varma2012, Jhajj2013,Zahedpour2014}. Impulsive Raman scattering induces a coherent superposition of rotational states $\vert J,m\rangle$ in N$_2$ and O$_2$, i.e. rotational wave packets 

\begin{equation}
\vert \Psi_R(t)\rangle = \sum_{J,m}c_{J,m}\vert J,m\rangle e^{-iE_Rt/\hbar},
\end{equation}

where $J$ and m are the quantum numbers of the total and z-component of the rotational angular momentum and $E_R= hcB J(J+1)$ the rotational state energies of the rigid rotor Hamiltonian (with $B$ the rotational constant, $h$ the Planck constant and $c$ the speed of light). The instantaneous non-adiabatic laser excitation locks in phase all the Raman-excited rotational states at $t = 0$, and the resulting wave packet evolves as in Eq. (1). For $t > 0$, the states in the sum dephase with respect to each other and the wave packet spreads. However, rephasing occurs in the form of revivals at specific times, leading to full re-alignment at periodic intervals of $\Delta t = T_R  =1/2Bc$ ($8.36$\,ps in N$_2$ and $11.2$\,ps in O$_2$). These effects have been measured using Coulomb explosion imaging in very low density gases \cite{Huang2006} and, importantly for laser propagation applications, the effects persist at atmospheric pressure \cite{Zahedpour2014,Chen2007}. Due to the even integer distribution of the rotational frequencies, additional alignment or anti-alignment recurrences also occur at $t =T_R/4$, $t =T_R/2$, and $t = 3T_R/4$ \cite{Huang2006,Chen2007}.\par

\begin{figure}[htbp!]
\centering\includegraphics[width=\textwidth]{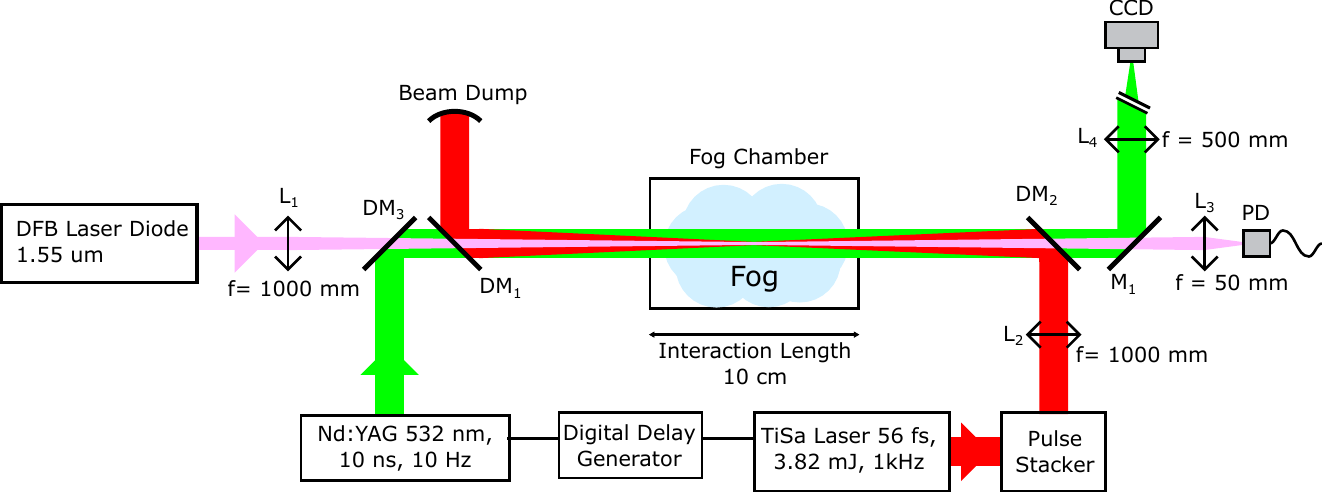}
\captionof{figure}{Quantum wake clearing of fog: a series of $8$ pulses of $56$\,fs are generated by a nested Michelson interferometer (pulse stacker \cite{Siders1998}) and softly focused into a fog chamber. The interval between the pulses can be tuned on- or off-resonance with the rotational revival times $T_R$ of the nitrogen molecules in air. Clearing of the fog is assessed with a counter-propagating CW laser at telecom wavelength ($1.55$\,$\mu$m) and a photodiode that measures the transmission. In addition a synchronized Nd:YAG laser is used for imaging the induced shockwave by shadowgraphy and interferometry}
\label{fig:yag}
\end{figure}

The periodic field-free alignment takes place until decoherence occurs, due to inter-molecular collisions and/or centrifugal distortion (deviation from the rigid rotor Hamiltonian). Typical rotational coherence times observed in N$_2$ at standard temperature and pressure are $\sim 80$\,ps \cite{Cryan2009}. Within this coherence time window, the rotational quantum wake can be coherently controlled by a train of short pulses precisely separated by the full revival time $T_R$, matching the \guillemotleft quantum echoes\guillemotright\,of the rotating molecule, as demonstrated by Cryan et al \cite{Cryan2009}. This coherent control method based on repetitive impulsive Raman excitation was originally proposed by Weiner et al \cite{Weiner1990} for controlling phonons in organic solids. By shaping the laser excitation as a sequence of $4$ pulses separated by $T_R$, Zahedpour et al \cite{Zahedpour2014} recently demonstrated that the molecular rotation of N$_2$ was driven to equivalent temperatures as high as $450$\,K ($30$\,meV/molecule of additional heating). Conversely, rotation was coherently hindered by applying series of pulses resonant at times for which molecules are anti-aligned $(T_R/2)$. Collisional decoherence and thermalization of the rotationally excited molecules after $\sim 100$\,ps \cite{Cheng2013} generates an on-axis  pressure spike that drives a single-cycle acoustic wave. The amplitude of this wave was shown to exceed that previously observed in plasma filaments \cite{Zahedpour2014}. In the present paper, we demonstrate that this plasma-free acoustic wave can also efficiently eject water droplets out of the beam, and that the coherently controlled quantum wake creates clear channels in fog without the need for filamentation-based ionization.

\section{Experiment}

For the experimental demonstration of the concept, we used a train of 8 ultrashort pulses, which provides an optimal compromise between the number of successive Raman excitations and the onset of decoherence due to collisions and centrifugal distortion.\par

A sketch of the experimental set-up is shown in Fig. \ref{fig:yag}. The laser pump consists of either a single pulse or a train of eight $\lambda=800$\,nm, $56$\,fs Ti:Sapphire (TiSa) pulses produced with a Michelson interferometer \cite{Siders1998} (\guillemotleft pulse stacker\guillemotright). The beam is focused with an f/$80$ lens into the fog chamber. The FWHM of the vacuum beam waist is about $w_{FWHM}=25$\,$\mu$m and the femtosecond laser train is repeated at a frequency of $1$\,kHz. The energy per pulse in the train is $0.48$\,mJ. A continuous wave (CW) counter-propagating laser at $\lambda=1.55$\,$\mu$m, spatially overlapping the TiSa beam, is used for measuring changes in transmission through the fog, and for investigating the dynamics of the droplets expulsion by the acoustic wave. In addition, a $10$\,Hz frequency doubled Nd:YAG laser ($\lambda = 532$\,nm) is synchronized to the TiSa pump beam to probe the 2D gas density at variable delays by shadowgraphy and interferometry. 

In order to simulate realistic fog/cloud conditions, the size distribution of the droplets in the cloud chamber is centered on a radius of $5$\,$\mu$m (measured by a Grimm 1.109 particle sizer). The droplet number density is variable from $0$ to $1.5\times10^5$\,cm$^{-3}$ (1000 times more than in a cumulus cloud) \cite{DeLaCruz2016}. This variation in  N$_d$ corresponds to attenuation ranging from $0$ dB (no fog) to $-10$\,dB. The droplets were generated and dispersed by a piezo-driven ultrasonic source.\par

\begin{figure}[htbp!]
\centering
	\includegraphics[width=\textwidth]{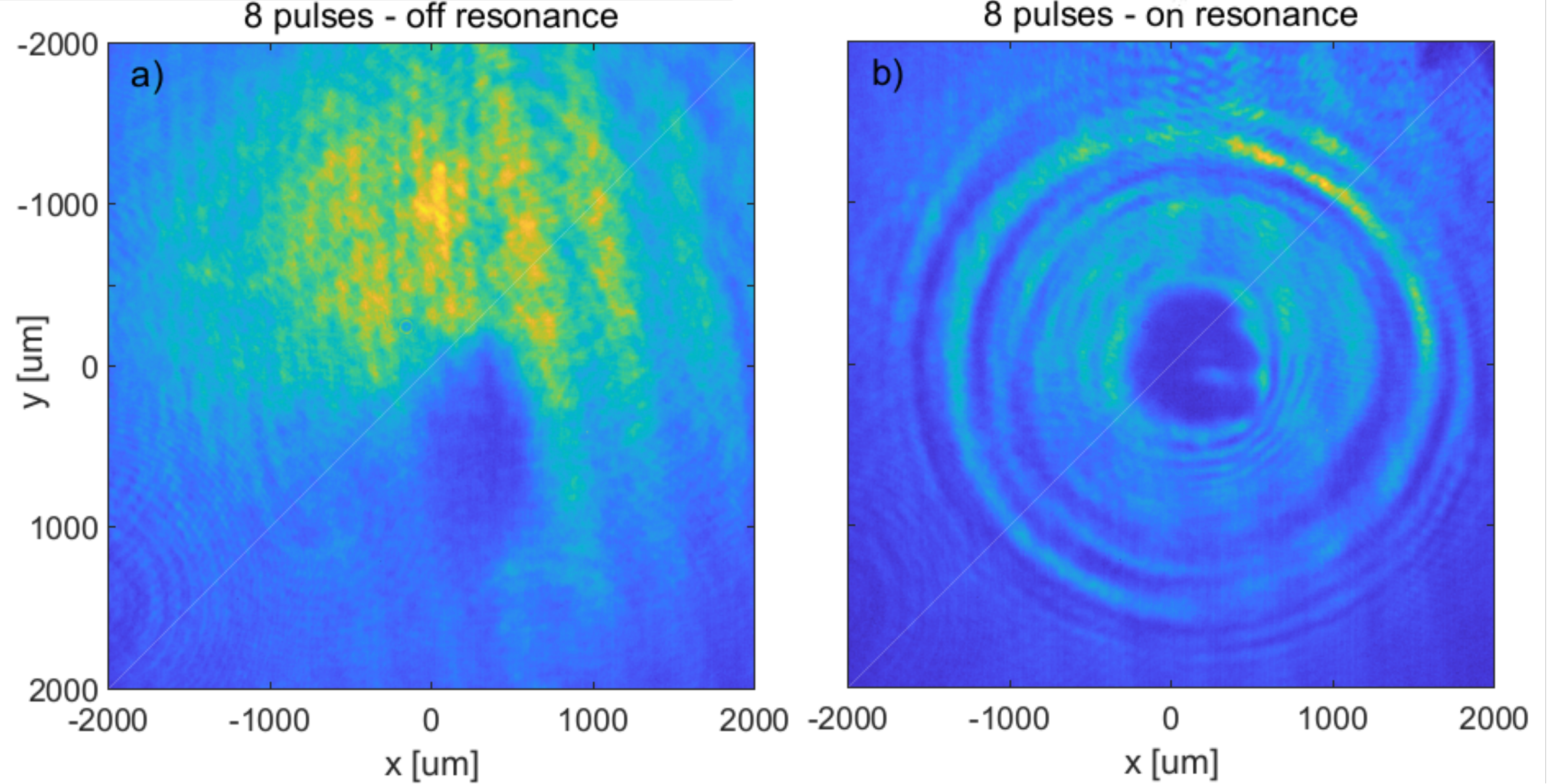}

\captionof{figure}{Coherent control of the rotational quantum wake induced shockwave in fog. Visualized through shadowgraphic measurements. Left: train with pulse intervals slightly detuned from resonance: $8.66$\,ps; Right: train of $8$ pulses tuned in resonance with the full revival time for N$_2$: $8.36$\,ps . Delay between TiSa pump and Nd:YAG probe: $4$\,$\mu$s. Total energy of the pump pulse train: $3.8$\,mJ}
	\label{fig:shadow}
\end{figure}

\begin{figure}[htbp]
\centering\includegraphics[width=\textwidth]{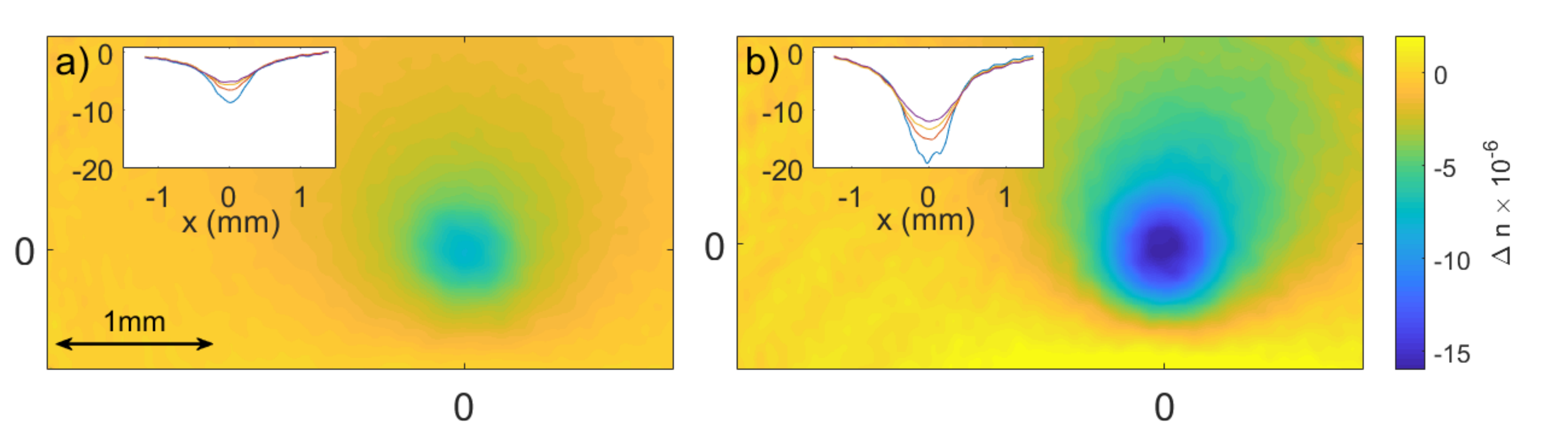}
\caption{Change in air refractive index due to rotational gas heating, measured by interferometry. Left: detuned pulse train: $8.66$\,ps; Right: resonant pulse train at full revival time for N$_2$: $8.36$\,ps . Delay between TiSa pump and Nd:YAG probe: $500$\,$\mu$s. The insets show the shift in the refractive index for time delays of $200$, $400$, $600$ and $800$ $\mu$s after the pump, in order of deepest to shallowest.}
\label{fig:density}
\end{figure}

\begin{figure}[htbp!]
\centering\includegraphics[width=\textwidth]{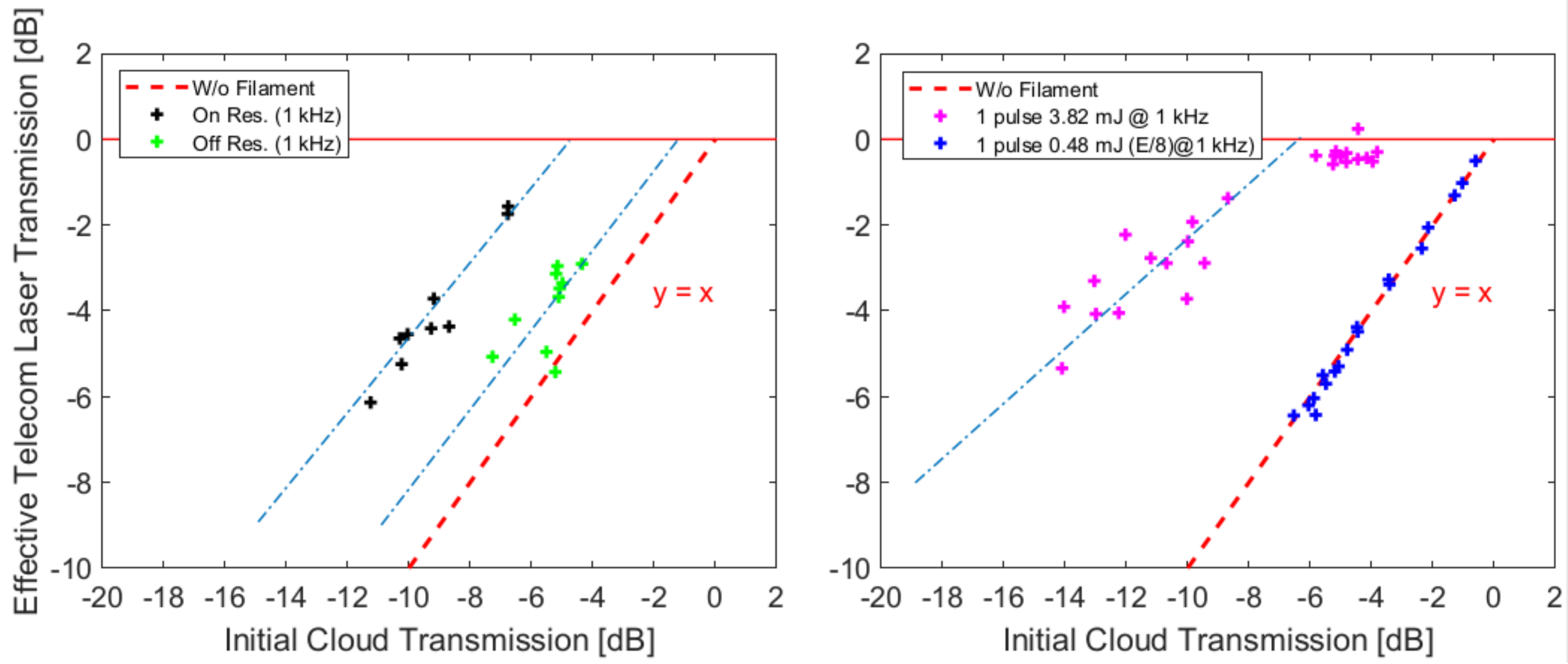}
\captionof{figure}{Fog clearing induced by the rotational quantum wake as a function of the initial signal attenuation caused by different fog densities 5(a). Clearing is efficient for all densities in the case of a pulse train in resonance with the rephasing time, and the efficiency increases with the initial density (see text). Clearing is strongly reduced for the off-resonant case. 5(b) comparison with the plasma induced shockwave - Purple: single pulse of $3.8$\,mJ (entire energy of the train in one pulse), Blue : single pulse of $0.48$\,mJ (one pulse of the train). Transmission is normalized to $1$ in each case, corresponding to the situation where no pump is present ($t < 0$).}
\label{fig:eff_plots}
\end{figure}

The shadowgraphic measurements allow for the observation of changes in the atmospheric pressure caused by the acoustic wave produced by the resonantly and off-resonantly spaced pulse trains. The shadowgraphic image in figure \ref{fig:shadow}(b) shows the acoustic wave produced in fog (droplet concentration $1.3\times10^5$\,cm$^{-3}$) when the pump laser consists of a train of $8$ pulses of $56$\,fs duration, separated by $8.36$\,ps, i.e. resonant with the full rephasing time of the rotational wave packet. Each of the $8$ pulses bears an intensity of the order of $\sim10^{13}$\,W/cm$^2$ at the beam waist. A clear radially expanding wave is produced, which reduces the air density in the center of the beam (typ. $500-700$\,$\mu$m diameter) during the first hundred $\mu$s after the pulse train. The reduced density region exhibits a lower refractive index (Fig. \ref{fig:density}(b)), which induces defocusing of the Nd:YAG probe laser, and thus a loss in the transmitted light. The depleted density region is eventually refilled by the surrounding gas after a few hundreds microseconds, as shown in the supplementary videos \textbf{Visualization 1} and \textbf{Visualization 2}. The acoustic wave expansion characteristics, in terms of amplitude and speed, appear almost unaffected by the presence of droplets, similar to previously reported measurements in clean air \cite{Wahlstrand2014}. Conversely, if the time separation between the pulses within the train is slightly detuned away from resonance ($8.66$\,ps), a much weaker shockwave is produced (Fig. \ref{fig:shadow}(a)), revealing a smaller temperature rise and refractive index shift in the gas (Fig. \ref{fig:density}(a)). These results demonstrate that the dominant mechanism responsible for the acoustic wave production here is the coherent control of the molecular quantum wake and not plasma generation (also confirmed by single pulse measurements, Fig. \ref{fig:eff_plots}).\par

A folded wavefront interferometer was insterted upstream of the imaging apparatus to measure the refractive index shift in the reduced density region surrounding the beam center. Gas density was extracted from retrieved interferograms using well-established fringe analysis techniques \cite{Takeda1982}. For these measurements no fog was added to the chamber, as to prevent turbulence in the fog from interfering with the acquisition of the interferograms. 
Figure \ref{fig:density}(a) shows the variation of the refractive index for the off-resonant case, where the pulses in the train are detuned by $300$\,fs relative to each other, at a pump-probe delay of $500$\,$\mu$s. The reduced density region is visible at the prior location of the beam center, gradually expanding outward from this point. At the center, the relative refractive index shift is $\Delta n_{\textit{off}}\approx-7.8\times10^{-6}$. If the pulse train is tuned to the revival time of molecular nitrogen, the reduced density region increases significantly in size and depth. At the same time delay of $500$\,$\mu$s, the resulting refractive index shift increases approximately by a factor of $2$ to $\Delta n_{\textit{on}}\approx-16.2\times10^{-6}$ (Fig. \ref{fig:density}(b)). The asymmetry visible in the figure stems from convection \cite{Jhajj2013}. In the insets the filling in of the reduced density region can be observed by the smoothing of the lineouts over time. The lineouts depict the spatially resolved change in refractive index along the x-axis, centered on the minimum value at the core.\par
These observations are consistent with the shadowgraphy measurements. The more efficient rotational molecular gas heating in the resonant case results in more heat deposition in the air. Consequently, the higher temperature gradient causes a stronger acoustic wave, which more efficiently reduces the air density over a more extended region.\par 

\begin{figure}[htbp!]
\centering\includegraphics[width=\textwidth]{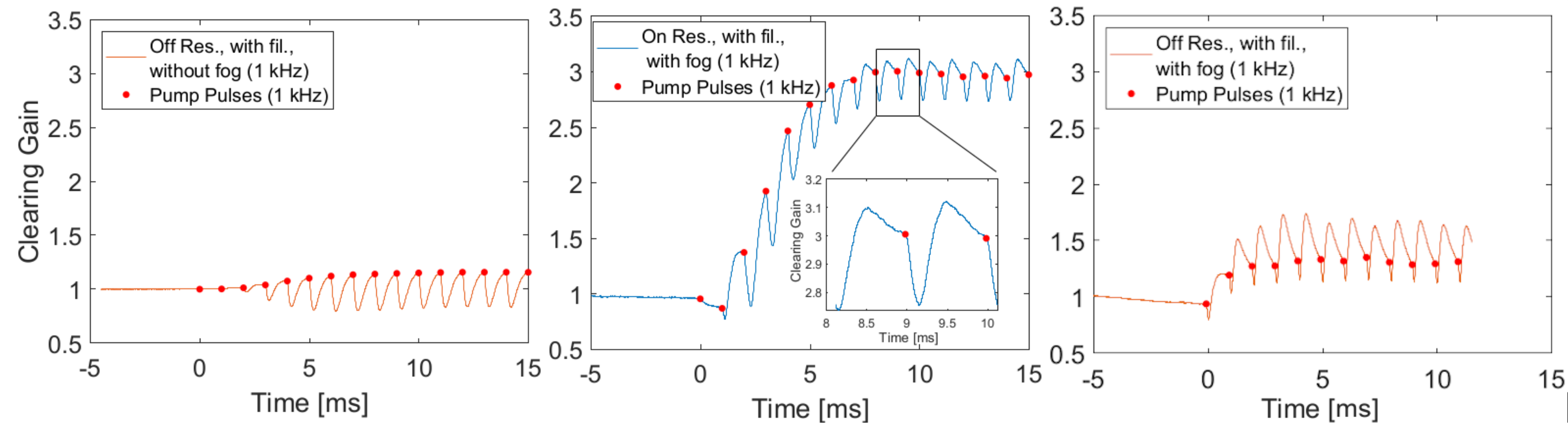}
\captionof{figure}{Fog clearing induced by the rotational quantum wake. Transmission gain in the case of (a) No fog with the $8$ pulse train pump laser, (b) Fog ($1.3\times10^5$\,cm$^{-3}$ droplet concentration) with the $8$ pulse train tuned on-resonance with the quantum wake rephasing time, and (c) Fog ($0.7\times10^5$\,cm$^{-3}$ droplet concentration) with the $8$ pulse train tuned off-resonance with the quantum wake rephasing time. Total energy of the pump: $3.8$\,mJ. Transmission is normalized to 1 in each case, corresponding to the situation where no pump is present ($t < 0$).}
\label{fig:deconv}
\end{figure}

The gain in transmission provided by the quantum wake clearing process was assessed by running the probe telecom laser in a CW mode (Fig. \ref{fig:deconv}). Fig. \ref{fig:deconv}(a) shows the effect of the 8-pulse train pump on the transmission of the $\lambda=1.55$\,$\mu$m CW laser in the absence of fog. The air density depletion due to rotational heating causes an initial defocusing of the telecom laser due to the induced refractive index gradient. This leads to the observed modulation of the signal at $1$\,kHz, due to the finite aperture of the detection system. In this geometry, the envelope of the modulation requires $2-3$\,ms to reach a steady state. In the presence of fog ($N_d\sim 1.3\times10^5$\,cm$^{-3}$), corresponding to $9$\,dB attenuation, the 8-pulse train dramatically increases the telecom laser transmission by a factor $3$ (i.e. $4.8$\,dB) when the pulse interval in the train is tuned on-resonance with the full revival time of $8.36$\,ps (Fig. \ref{fig:deconv}(b)). Interestingly, the modulated envelope reaches a steady state after $\sim 7-8$\,ms, longer than in the droplet-free case. This can be interpreted by the fact that the first few 8-pulse trains, delivered at $1$\,kHz, are used to clear the initial local droplet concentration to make its way for the following pulses. Conversely, when the 8-pulse interval is tuned slightly off-resonance, the gain drops to a factor of only $\sim1.5$ ($1.7$\,dB, Fig \ref{fig:deconv}(c)). This remaining gain is caused by partial rotational heating.\par
 
The efficiency of rotational heating-induced fog clearing was assessed as a function of the fog's optical density, by varying the droplet number density in the chamber. Although large fluctuations are present due to instabilities in the cloud chamber (entrance and exits are open to the ambient air), a clear enhancement in transmission is observed for all the optical density values in the case of the on-resonance pump pulse trains. As quantitative assessment of the process efficiency, we consider the maximum initial cloud extinction value for which transmission is fully recovered ($100\%$, i.e. $0$\,dB extinction). For the on-resonance pulse train, this value corresponds to $-6$\,dB initial extinction, while for the slightly detuned pulse train, it drops to $-2$\,dB. It is now relevant to compare these values to the values achieved with a single pulse (Fig. \ref{fig:eff_plots}(b) (with all other experimental conditions kept identical to the 8 pulses train case). When only one pulse of the train is used (bearing $0.48$\,mJ as in the pulse train), no transmission gain is observed, which suggests that the plasma generation of individual pulses of the train is negligible. Conversely if the whole energy of the pulse train, $3.8$\,mJ, is concentrated in a single pulse, plasma generation and filamentation are initiated, and we observe the associated acoustic wave clearing process seen previously \cite{Schimmel2018, DeLaCruz2016}, i.e. $-8$\,dB, which exceeds the quantum wake clearing process with 8 pulses. Also, the slope significantly exceeds the slope in the case of rotational heating with the pulse train. The linear increase in the transmission gain was previously attributed \cite{Schimmel2018} to the fact that the number of droplets $\Delta N$ ejected by the shockwave is proportional to the number of droplets $N$ initially present in the volume (percentage of droplets cleared in the channel: on resonance: $57.4\%$; off resonant: $28.0\%$ ; 1 pulse full energy : $75.1\%$ ; 1 pulse 1/8th of energy: $0\%$). In other words, the energy of the shockwave is not significantly altered by the quantity of droplets that have to be moved. \par

As already mentioned above, the clear enhancement of the transmission at the rephasing time of the N$_2$ rotational wave packet, as compared to the non-resonant case, excludes dominant clearing processes like droplet explosion, shattering or direct evaporation by the laser beam. However, in the present configuration, the rotational heating of the gas itself ($450$\,K) exceeds the vaporization temperature of water, so that the partial or complete evaporation of the droplets by heat transfer from the hot air to the droplets may have a contribution in the fog clearing as well. Although the exact calculation of the transient evaporated mass rate from a moving droplet is a complex problem \cite{Sirignano2010}, some qualitative arguments can be drawn for assessing this role in the clearing. A necessary (but not sufficient) condition for the process to occur is that the amount of energy $Q$ required for vaporizing the ensemble of droplets in the channel is less than or equal to the energy deposited by the laser. With $1.5\times10^5$\,cm$^{-3}$ droplets in a channel of $0.5$\,mm radius and $10$\,cm length, this leads to $Q = 13$\,mJ, which means that all the incoming laser energy should be deposited in the gas and then transferred from the heated gas to the droplets with an efficiency of $100\%$. These considerations suggest that the droplet vaporization by the heated gas does not play the major role in the fog clearing process, in contrast to the displacement of the drops by the acoustic wave as previously observed for the filament induced clearing process \cite{Schimmel2018, DeLaCruz2016, Wolf2018}. 

\section{Conclusion}

In this article, we demonstrated that neither ionization nor filamentation are required to produce a fog clearing acoustic wave, as they can be substituted by quantum controlled rotational heating in air. This discovery opens new perspectives in the application of the method for real, atmospheric scale operation like FSO. For instance, intense mid-IR lasers, which provide only moderate ionization as they propagate in air \cite{Mongin2016, Mitrofanov2015, Liang2016, Kartashov2013}, are now potentially very attractive candidates for fog clearing applications. It was recently reported in particular that TW peak powers CO$_2$ lasers can provide wide diameter self-guiding channels over kilometric distances \cite{Tochitsky2019}, fulfilling the requirements for FSO between earth and satellites.

\section{Funding}
Schweizerischer Nationalfonds zur F\"orderung der Wissenschaftlichen Forschung (SNF) (200021-178926); Horizon 2020 Framework Programme (H2020) (737033-LLR); Office of Naval Research
(ONR) (N00014-17-1-2705, N00014-17-1-2778); Air Force Office of Scientific Research (AFOSR) (FA9550-16-1-0284, FA9550-16-1-0121).

\section{Acknowledgments}

MCS, TP and JPW gratefully acknowledge technical support from Michel Moret and financial support from the Schweizerischer Nationalfonds zur F\"orderung der Wissenschaftlichen Forschung under grant no 200021-178926. TP and JPW acknowledge the European Union's Horizon 2020 research and innovation program under grant agreement no 737033-LLR. \par 
IL, EWR, and HMM gratefully acknowledge support from the Office of Naval Research
(ONR) (N00014-17-1-2705, N00014-17-1-2778) and the Air Force Office of Scientific Research (AFOSR) (FA9550-16-1-0284, FA9550-16-1-0121).

\section{Disclosures}

The authors declare no conflicts of interest.

\bibliography{bib}

\end{document}